\documentclass[twocolumn,showpacs,preprintnumbers,amsmath,amssymb]{revtex4}
\usepackage{graphicx}% Include figure files
\usepackage{epstopdf}
\usepackage{dcolumn}% Align table columns on decimal point
\usepackage{bm}% bold math
\begin{document}

\preprint{}
\title{A limit on Short-range Modifications to Gravity }
\author{P. J. S. Watson}
\email{watson@physics.carleton.ca}
\affiliation{Ottawa-Carleton Institute for Physics, Carleton University, \\ Ottawa, Canada, K1S 5B6
}

\date{\today}

\begin{abstract}
There has been considerable interest in providing new limits on the short range behaviour of gravity, or, in general, anomalous short-range interactions. In this note we show that one use the interaction of ultra-cold neutrons to obtain a better limit below about 10 nm.
 \end{abstract}
\maketitle

\section{Introduction}
There has been a great deal of interest lately in theories  in which the gravitational interaction is modified at distances which are large compared to the weak scale \cite{Arkani-Hamed:1998rs},\cite{Hewett:2002hv}. Conventional gravitational experiments put very good limits on any corrections to Newtonian gravity at distances $d>\sim 150\mu $ \cite{Hoyle:2000cv} and recent experiments \cite{Chiaverini:2002cb} have shown that there are no gravity-like forces with interaction strengths $G'>10^3G$ for $d>\sim 10\mu $. There are weak limits  below $\sim 10\mu $ \cite{Fischbach:2001ry}, and there are various suggestions  \cite{Dimopoulos:2003mw}, \cite{Watson:2003dq} as to how these could be improved. Here we point out that neutron scattering serves  to provide a limit which extends well below the $\sim 10\mu $ range. 

We can assume that the anomalous interaction is given by
% MathType!MTEF!2!1!+-
% faaafaart1ev1aaat0uyJj1BTfMBaerbuLwBLnharmWu51MyVXgaru
% WqVvNCPvMCaebbnrfifHhDYfgasaacH8srps0lbbf9q8WrFfeuY-Hh
% bbf9v8qqaqFr0xc9pk0xbba9q8WqFfea0-yr0RYxir-Jbba9q8aq0-
% yq-He9q8qqQ8frFve9Fve9Ff0dmeaabaqaciGacaGaaeqabaWaaeaa
% eaaakeaacaWGwbWaaeWaaeaacaWGYbaacaGLOaGaayzkaaGaeyypa0
% JaeyOeI0Iaam4zamaaCaaaleqabaGaaGOmaaaakiabl+qiOjaadoga
% daWcaaqaaiaadwgadaahaaWcbeqaaiabgkHiTiaadkhacaGGVaGaeq
% 4UdWgaaaGcbaGaamOCaaaacqGH9aqpcqGHsislcqaHXoqycaWGhbGa
% amyBamaaBaaaleaacaaIXaaabeaakiaad2gadaWgaaWcbaGaaGOmaa
% qabaGcdaWcaaqaaiaadwgadaahaaWcbeqaaiabgkHiTiaadkhacaGG
% VaGaeq4UdWgaaaGcbaGaamOCaaaaaaa!4AFC!
\begin{equation} \label{V_a} 
V\left( r \right) =  - g^2 \hbar c\frac{{e^{ - r/\lambda } }}{r} =  - \alpha Gm_1 m_2 \frac{{e^{ - r/\lambda } }}{r}
\end{equation}

The first expression represent a generic fifth force, while the second explicitly assumes an interaction which couples to mass. In what follows, we choose the second form. We adopt the notation of Fischbach et al. \cite{Fischbach:2001ry}.

\section{Neutron Scattering}
 Ultra-cold neutrons (UCN's) \cite{golub} have energies $E<10^{-6}eV$ and correspondingly long wavelengths, so the neutron scatters coherently off the atomic nuclei in a lattice. 
It is well known  \cite{golub},  \cite{Steyerl} that there is a bulk interaction of the neutron with any material given by the coherent interaction of the neutron with all the nuclei. This  gives rise to a potential of the form 
% MathType!MTEF!2!1!+-
% faaafaart1ev1aqat0uyJj1BTfMBaerbuLwBLnharmWu51MyVXgaru
% WqVvNCPvMCaebbnrfifHhDYfgasaacH8srps0lbbf9q8WrFfeuY-Hh
% bbf9v8qqaqFr0xc9pk0xbba9q8WqFfea0-yr0RYxir-Jbba9q8aq0-
% yq-He9q8qqQ8frFve9Fve9Ff0dmeaabaqaciGacaGaaeqabaWaaeaa
% eaaakeaacaWGvbWaaSbaaSqaaiaadohaaeqaaOGaeyypa0ZaaSaaae
% aacaaIYaGaeyiWdaNaeS4dHG2aaWbaaSqabeaacaaIYaaaaOGaamOt
% aiaadkgaaeaacaWGTbWaaSbaaSqaaiaad6eaaeqaaaaaaaa!3814!
\begin{equation}
 \label{Us} U_s  = \frac{{2\pi \hbar ^2 Nb}}{{m_N }}
\end{equation}
where b is the scattering length, N is the number density of nuclei and ${m_N }$ the neutron mass.This gives rise to a critical velocity $v_c$ below which neutrons are totally reflected from the surface.
For Be the relevant parameters are %MathType!ZZhx47!eaaaduGcbiqGIjqG9iqG3iqGUiqG4iqGGiqGMjqGTjqGSiqGGiqGbj
%qG9iqG5iqGSiqGGiqGojqG9iqG3iqGUiqG4iGH0kaIXiaIWmWdaaWc
%reGaikJaiIdaaOGacYIedw1apeaaleGedoharaGccyypcGymciOlcG
%OncyiTcGymcGimd8aaaSqebiGHTiaI3aaakiXGLjXGwbaa!1F74!
$b=7.8\ fm, \ N=1.24\times 10^{28},U_s=2.5\times 10^{-7}eV,v_c=6.9ms^{-1}$. 
However, the anomalous interaction (\ref{V_a}) gives rise to an additional uniform potential inside a solid given by
% MathType!MTEF!2!1!+-
% faaafaart1ev1aqat0uyJj1BTfMBaerbuLwBLnharmWu51MyVXgaru
% WqVvNCPvMCaebbnrfifHhDYfgasaacH8srps0lbbf9q8WrFfeuY-Hh
% bbf9v8qqaqFr0xc9pk0xbba9q8WqFfea0-yr0RYxir-Jbba9q8aq0-
% yq-He9q8qqQ8frFve9Fve9Ff0dmeaabaqaciGacaGaaeqabaWaaeaa
% eaaakeaacaWGwbWaaSbaaSqaaiaadofaaeqaaOGaeyypa0JaeyOeI0
% IaaGinaiabgc8aWjabeg7aHjaadEeacaWGTbWaaSbaaSqaaiaad6ea
% aeqaaOGaeqyWdi3aa8qmaeaadaWcaaqaaiaadwgadaahaaWcbeqaai
% abgkHiTiaadkhacaGGVaGaeq4UdWgaaaGcbaGaamOCaaaacaWGYbWa
% aWbaaSqabeaacaaIYaaaaOGaamizaiaadkhacqGH9aqpaSqaaiaaic
% daaeaacqGHEisPa0Gaey4kIipakiabgkHiTiaaisdacqGHapaCcqaH
% XoqycaWGhbGaamyBamaaBaaaleaacaWGobaabeaakiabeg8aYjabeU
% 7aSnaaCaaaleqabaGaaGOmaaaaaaa!5501!
\begin{equation} \label{interact}
V_s  =  - 4\pi \alpha Gm_N \rho \int_0^\infty  {\frac{{e^{ - r/\lambda } }}{r}r^2 dr = }  - 4\pi \alpha Gm_N \rho \lambda ^2 \end{equation}

where  $\rho$ is the mass density. This provides a universal attractive potential between any material and the neutron: it is roughly equivalent to a work-function.

Since a neutron bottle can be constructed, obviously
%MathType!ZZhx47!eaaaduGcbiXGvnWdbaWcbiXGZbqeaOGag2IedA1apeaaleGedohara
%GccyOpcGimaaa!06A5!
$U_s+V_s>0$. This gives rise to a limit on $\alpha$ of the form
% MathType!MTEF!2!1!+-
% faaafaart1ev1aaat0uyJj1BTfMBaerbuLwBLnharmWu51MyVXgaru
% WqVvNCPvMCaebbnrfifHhDYfgasaacH8srps0lbbf9q8WrFfeuY-Hh
% bbf9v8qqaqFr0xc9pk0xbba9q8WqFfea0-yr0RYxir-Jbba9q8aq0-
% yq-He9q8qqQ8frFve9Fve9Ff0dmeaabaqaciGacaGaaeqabaWaaeaa
% eaaakeaacqaHXoqycqGHKjYOdaWcaaqaaiaadggadaWgaaWcbaGaam
% yqaaqabaaakeaacqaH7oaBdaahaaWcbeqaaiaaikdaaaaaaOGaaiil
% aiaadggadaWgaaWcbaGaamyqaaqabaGccqGH9aqpdaWcaaqaaiabl+
% qiOnaaCaaaleqabaGaaGOmaaaakiaad6eacaWGIbaabaGaaGOmaiaa
% d2gadaqhaaWcbaGaamOtaaqaaiaaikdaaaGccqaHbpGCcaWGhbaaai
% abg2da9maalaaabaGaeS4dHG2aaWbaaSqabeaacaaIYaaaaOGaamOy
% aaqaaiaaikdacaWGTbWaa0baaSqaaiaad6eaaeaacaaIYaaaaOGaam
% yBamaaBaaaleaacaWGbbaabeaakiaadEeaaaaaaa!4C65!
\begin{equation} \label{Be}
\alpha  \le \frac{{a_A }}{{\lambda ^2 }},a_A  = \frac{{\hbar ^2 Nb}}{{2m_N^2 \rho G}} = \frac{{\hbar ^2 b}}{{2m_N^2 m_A G}}
\end{equation}
where ${m_A }$ is the nuclear mass. 
A crude limit comes from regarding this as an equality: for Be, we find $a_{Be}   \approx 1.5 \times 10^7 m^2 $

Obviously this can be improved by using a heavy nucleus with a small scattering length. It appears that Pb would be optimal, but there does not appear to be any experiment on the reflection of UCN's  from Pb. However, if an actual measurement  of the critical velocity ${v_c }$ could be made, the limit is then given by the parameter 
% MathType!MTEF!2!1!+-
% faaafaart1ev1aaat0uyJj1BTfMBaerbuLwBLnharmWu51MyVXgaru
% WqVvNCPvMCaebbnrfifHhDYfgasaacH8srps0lbbf9q8WrFfeuY-Hh
% bbf9v8qqaqFr0xc9pk0xbba9q8WqFfea0-yr0RYxir-Jbba9q8aq0-
% yq-He9q8qqQ8frFve9Fve9Ff0dmeaabaqaciGacaGaaeqabaWaaeaa
% eaaakeaacaWGHbWaa0baaSqaaiaadgeaaeaacaGGNaaaaOGaeyypa0
% ZaaSaaaeaacqWIpecAdaahaaWcbeqaaiaaikdaaaGccaWGIbaabaGa
% amyBamaaDaaaleaacaWGobaabaGaaGOmaaaakiaad2gadaWgaaWcba
% GaamyqaaqabaGccaWGhbaaamaalaaabaGaeqiTdqMaamODaaqaaiaa
% dAhadaWgaaWcbaGaam4yaaqabaaaaaaa!3D8B!
\begin{equation} \label{Pb}
a_A' = \frac{{\hbar ^2 b}}{{m_N^2 m_A G}}\frac{{\delta v}}{{v_c }}
\end{equation}
If (say) the critical velocity for Pb could be measured to 
$\frac{{\delta v}}{{v_c }} = 1\% $
, this would give 
$a_{Pb}'  \approx 1.6 \times 10^4 m^2 $, improving the limit by about 1000.
Below we show the limits on  $\alpha $ as a function of $\lambda $ that could be obtained from this method, comparing it with various other limits. In principle, this could provide a limit on strong, gravity-like forces down to a distance of 0.01 nm.

\begin{figure}[htbp]
\includegraphics[width=3in]{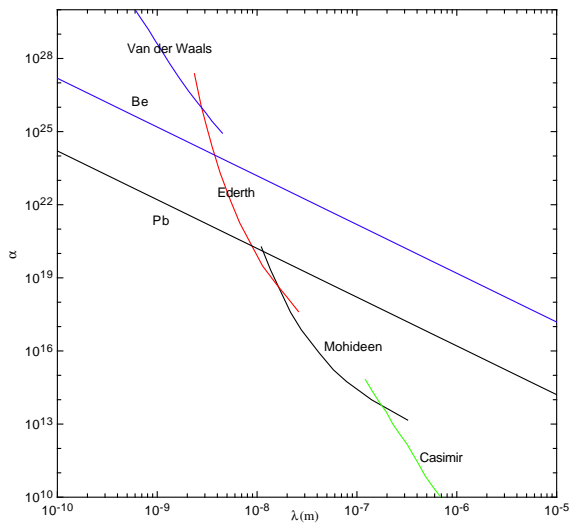}
\caption{Limits on $\alpha $ as a function of $\lambda $. The straight line labelled "Be" show the actual limit from (\ref{Be}). The line labeled "Pb" shows the limit that could in principle be obtained from (\ref{Pb}). The remaining lines are taken from Fischbach et al. \cite{Fischbach:2001ry}.}
\label{fig1}
\end{figure}

There have also been suggestions that the new interaction may be of a power law form. If this is used in (\ref{interact}), the integral is divergent unless a short distance cutoff is applied. The value of this is quite arbitrary, and hence one cannot provide a useful limit.

A related effect of $V_s$ would be to modify the refractive index of UCN's in a material. In the case where the scattering length is negative, this would in turn would modify the 1/v law for neutron absorption \cite{Robson}, since effectively the velocity of neutrons would be higher than anticipated, and hence the absorption cross-section would be decreased. It is amusing that just such an effect is seen in Gd \cite{Rauch} however it is extremely unlikely that the effect there is due to an anomalous long-range interaction.
 
\begin{acknowledgments} This work was supported by NSERC. I am grateful to Richard Hemingway for commenting on the manuscript.
\end{acknowledgments}

 \end{document}